\documentclass[journal]{IEEEtran}
\ifCLASSINFOpdf
   \usepackage[pdftex]{graphicx}
\else
\fi
\usepackage{amsmath}
\usepackage{array}
\ifCLASSOPTIONcompsoc
  \usepackage[caption=false,font=normalsize,labelfont=sf,textfont=sf]{subfig}
\else
  \usepackage[caption=false,font=footnotesize]{subfig}
\fi
\usepackage{fixltx2e}
\usepackage{stfloats}

\usepackage{csvsimple}
\usepackage{color}
\usepackage{hyperref}

\DeclareMathOperator{\Enc}{Enc}
\DeclareMathOperator{\Dec}{Dec}

\DeclareMathOperator*{\argmin}{arg\,min}

\begin{document}

\title{Autoencoders for Unsupervised Anomaly Segmentation in Brain MR Images: A Comparative Study}

%\author{Christoph~Baur,~\IEEEmembership{Member,~IEEE,}
%        Stefan~Denner,~\IEEEmembership{Fellow,~OSA,}
%        Benedikt~Wiestler,~\IEEEmembership{Fellow,~OSA,}
%        Shadi~Albarqouni,~\IEEEmembership{Fellow,~OSA,}
%        and~Nassir~Navab,~\IEEEmembership{Life~Fellow,~IEEE}% <-this % stops a space
\author{Christoph~Baur,
        Stefan~Denner,
        Benedikt~Wiestler,
        Shadi~Albarqouni
        and~Nassir~Navab
\thanks{C. Baur, S. Denner, S. Albarqouni, N. Navab are with the Chair for Computer Aided Medical Procedures (CAMP), TU Munich, Boltzmannstr. 3, Garching near Munich}% <-this % stops a space
\thanks{S. Albarqouni is with the Computer Vision Laboratory, ETH Zurich, Sternwartstrasse 7,
Zurich, Switzerland}% <-this % stops a space
\thanks{B. Wiestler is with the Neuroradiology Department of Klinikum Rechts der Isar, Ismaningerstr. 22, Munich, Germany}% <-this % stops a space
\thanks{N. Navab is with the Whiting School of Engineering, Johns Hopkins University, Baltimore, United States}}% <-this % stops a space


\markboth{Journal of \LaTeX\ Class Files,~Vol.~14, No.~8, August~2015}%
{Shell \MakeLowercase{\textit{et al.}}: Bare Demo of IEEEtran.cls for IEEE Journals}



\maketitle

\begin{abstract}
% V1
Deep unsupervised representation learning has recently led to new approaches in the field of Unsupervised Anomaly Detection (UAD) in brain MRI. The main principle behind these works is to learn a model of normal anatomy by learning to compress and recover healthy data. This allows to spot abnormal structures from erroneous recoveries of compressed, potentially anomalous samples. The concept is of great interest to the medical image analysis community as it i) relieves from the need of vast amounts of manually segmented training data---a necessity for and pitfall of current supervised Deep Learning---and ii) theoretically allows to detect arbitrary, even rare pathologies which supervised approaches might fail to find. To date, the experimental design of most works hinders a valid comparison, because i) they are evaluated against different datasets and different pathologies, ii) use different image resolutions and iii) different model architectures with varying complexity. The intent of this work is to establish comparability among recent methods by utilizing a single architecture, a single resolution and the same dataset(s). Besides providing a ranking of the methods, we also try to answer questions like i) how many healthy training subjects are needed to model normality and ii) if the reviewed approaches are also sensitive to domain shift. Further, we identify open challenges and provide suggestions for future community efforts and research directions.

\end{abstract}


% \fi
\IEEEpeerreviewmaketitle

\section{Introduction}
%\IEEEPARstart{T}{his} demo file is intended to serve as a ``starter file''
%for IEEE journal papers produced under \LaTeX\ using
%IEEEtran.cls version 1.8b and later.
%I wish you the best of success.

% Snippets
%Snippet from \cite{you2019uad}
%  lesion detection has taken huge strides in automated lesion detection of prespecified type, where models are optimized to detect lesions contained in a training set(Ayachi and Amor, 2009; Zikic et al., 2012; Geremia et al., 2011; Dong et al., 2017; Pereira et al., 2016; Kamnitsas et al., 2017; Li et al., 2018)

\IEEEPARstart{M}{R} imaging of the brain is at the heart of diagnosis and treatment of neurological diseases. When sifting MR scans, Radiologists intuitively rely on a learned model of normal brain anatomy to detect pathologies. However, reading and interpreting MR scans is an intricate process: It is estimated that in 5-10\% of scans, a relevant pathology is missed~\cite{bruno2015understanding}. Recent breakthroughs in machine learning have led to automated medical image analysis methods which achieve great levels of performance in the detection of tumors or lesions arising from neuro-degenerative diseases such as Alzheimers or Multiple Sclerosis (MS). Despite all their outstanding performances, these methods---mainly based on Supervised Deep Learning---carry some disadvantages: 1) their training calls for large and diverse annotated datasets, which are scarce and costly to obtain; 2) the resulting models are limited to the discovery of lesions which are similar to those in the training data. This is especially crucial for rare diseases, for which collecting training data poses a great challenge. Lately, there have been some Deep Learning-driven attempts towards automatic brain pathology detection which tackle the problem from the perspective of so-called Unsupervised Anomaly Detection (UAD). These approaches are more similar to how Radiologists read MR scans, do not require data with pixel-level annotations and have the potential to detect arbitrary anomalies without a-priori knowing about their appearances.

% From the Workshop paper
%which can be grouped into methods based on statistical modeling, content-based retrieval or clustering and outlier detection~\cite{TaboadaCrispi:te}. Weiss et al.~\cite{Weiss:2013cb} employed Dictionary Learning and Sparse Coding to learn a representation of normal brain patches in order to detect MS lesions. Other unsupervised MS lesion segmentation methods rely on thresholding and 3D connected component analysis~\cite{Iheme:2013hj} or fuzzy c-means clustering with topology constraints~\cite{Shiee:2010dw}.

% From the workshop paper
% A plethora of work in the field of deep learning based UAD has been devoted to videos primarily based on Autoencoders (AEs) due to their ability to express non-linear transformations and the ability to detect anomalies directly from poor reconstructions of input data\cite{Sabokrou:2016gf,Hasan:2016gb,Chong:2017vb}

UAD has a long history in medical image analysis and in brain imaging in particular. Traditional methods are based on statistical modeling, content-based retrieval, clustering or outlier-detection. A review on such classical approaches with a focus on brain CT imaging is given in \cite{taboada2009anomaly}.
Since the rise of Deep Learning, a plethora of new, data-driven approaches has appeared. Initially, Autoencoders (AEs), with their ability to learn non-linear transformations of data onto a low-dimensional manifold, have been leveraged for cluster-based anomaly detection. Lately, a variety of works used AEs and generative modeling to not simply detect, but localize and segment anomalies directly in image-space from imperfect reconstructions of input images, which is surveyed in this work in the context of brain MRI.
%For a comparison of classical and Deep Learning based approaches for anomaly segmentation at the example of street images, the interested reader is refered to \cite{hendrycks2019benchmark}.

The underlying idea thereby is to model the distribution of healthy anatomy of the human brain with the help of deep (generative) representation learning. Once trained, anomalies can be detected as outliers from the modeled, normative distribution. AEs~\cite{baur2018deep}\cite{atlason2019unsupervised} and their generative siblings~\cite{baur2018deep}\cite{chen2018unsupervised}\cite{zimmerer2019unsupervised}\cite{zimmerer2018context}\cite{pawlowski2018unsupervised} have emerged as a popular framework to achieve this by essentially learning to compress and reconstruct MR data of healthy anatomy. The respective methods can essentially be divided into two categories: 1) Reconstruction-based approaches compute a pixel-wise discrepancy between input samples and their feed-forward reconstructions to determine anomalous lesions directly in image-space; 2) Restoration-based methods~\cite{schlegl2017unsupervised}\cite{you2019uad} try to alter an input image by moving along the latent manifold until a normal counterpart to the input sample is found, which in turn is used again to detect lesions from the pixel-wise discrepancy of the input data and its healthy restoration. To date---albeit all of these methods report promising performances---results can hardly be compared and drawing general conclusions on their strengths \& weaknesses is barely possible. This is hindered by the following issues: i) most of the works rely on very different datasets with barely overlapping characteristics for their evaluation, ii) are evaluated against different pathologies, iii) operate on different resolutions and iv) utilize different model architectures with varying model complexity. The main intent of this work is to establish comparability among a broad selection of recent methods by utilizing---where applicable---a single network architecture, a single resolution and the same dataset(s).

\textbf{Contribution}|Here, we provide a comparative study of recent Deep-Learning based UAD approaches for brain MRI. We compare various reconstruction- as well as restoration based methods against each other on a variety of different MR datasets with different pathologies\footnote{Code will be made publicly available at \url{https://github.com/StefanDenn3r/unsupervised_anomaly_detection_brain_mri} after successful peer-review of the manuscript.}. The models are tested on four different datasets for detecting two different pathologies. To evaluate the methods without having to make general assumptions about what constitutes a detection, we utilize pixel-wise segmentation measures as a tight proxy for UAD performance. For a fair comparison, we determined a single, unified architecture on which all the methods rely in this study. This ensures that model complexity is the same for all approaches, if applicable. The performances of the originally proposed networks are also presented. Further, we provide insights on the number of healthy training samples and their impact on model performance, and peek at generalization capabilities of AE models. %and briefly look into epistemic uncertainty of the trained networks.

\section{Unsupervised Deep Representation Learning for Anomaly Detection}
\subsection{Modeling Healthy Anatomy}

\begin{figure*}[!t]
\includegraphics[width=\textwidth]{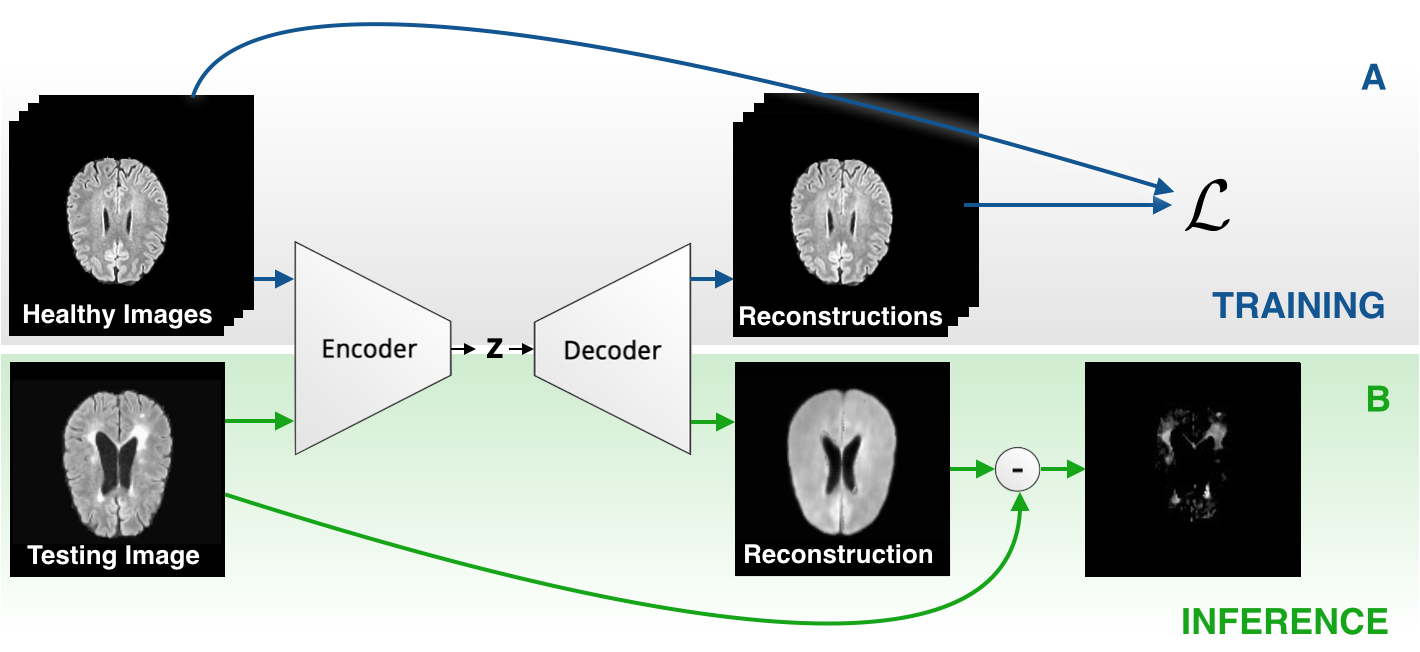}
\caption{The concept of Autoencoder-based Anomaly Detection/Segmentation: A) Training a model from only healthy samples and B) anomaly segmentation from erroneous reconstructions of input samples, which might carry an anomaly.}
\label{fig:concept}
\end{figure*}

\begin{figure*}[!t]
\centering
\subfloat[AE]{\includegraphics[width=2.5in]{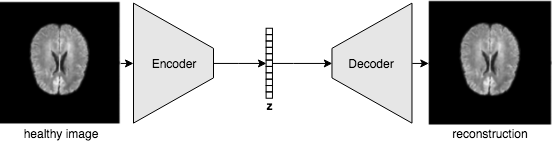}%
\label{fig:ae}}
\hfil
\subfloat[VAE]{\includegraphics[width=2.5in]{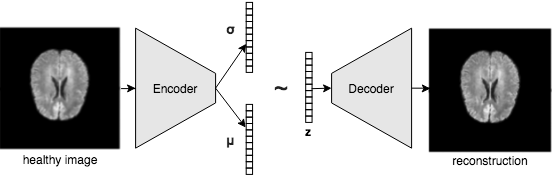}%
\label{fig:vae}}

\subfloat[AAE]{\includegraphics[width=2.5in]{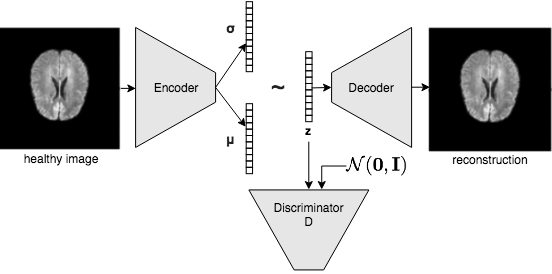}%
\label{fig:aae}}
\hfil
\subfloat[AnoVAEGAN]{\includegraphics[width=2.5in]{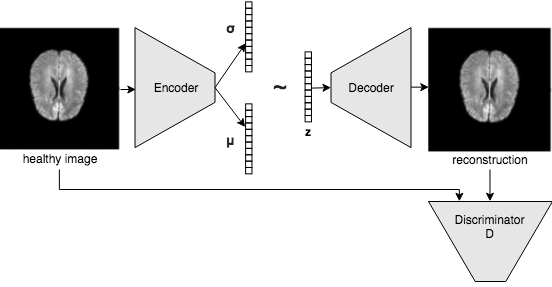}%
\label{fig:anovaegan}}

\subfloat[Context AE]{\includegraphics[width=2.5in]{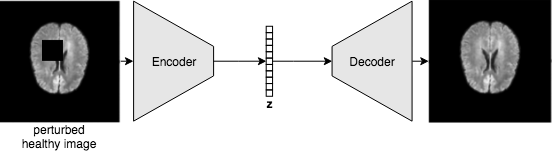}%
\label{fig:contextae}}
\hfil
\subfloat[GAN]{\includegraphics[width=2.5in]{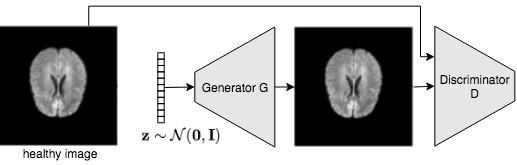}%
\label{fig:gan}}

\caption{Autoencoder-based architectures for UAD at a glance}
\label{fig:ae_architectures}
\end{figure*}

The core concept behind the reviewed methods is the modeling of healthy anatomy with unsupervised deep (generative) representation learning. Therefor, the methods leverage a set of healthy MRI scans $\mathcal{X}_{healthy} \in \mathcal{R}^{D\times H \times W}$ and learn to project it to and recover it from a lower dimensional distribution $\mathbf{z} \in \mathcal{R}^K$ (see Fig. \ref{fig:concept}). In the following, we first shed the light on the ways how this normative distribution can be modeled, and then present different approaches how anomalies can be discovered using trained models.

\textbf{Autoencoders}|Early work in this field relied on classic AEs (Fig. \ref{fig:ae}) to model the normative distribution: An encoder network $\Enc_{\theta}(\mathbf{x})$ with parameters $\theta$ is trained to project a healthy input sample $\mathbf{x} \in \mathcal{X}_{healthy}$ to a lower dimensional manifold $\mathbf{z}$, from which a decoder $\Dec_{\phi}(\mathbf{z})$ with parameters $\phi$ then tries to reconstruct the input as $\mathbf{\hat{x}} = \Dec_{\phi}(\Enc_{\theta}(\mathbf{x}))$. In other words, the model is trained to compress and reconstruct healthy anatomy by minimizing a reconstruction loss $\mathcal{L}$

\begin{equation}
    \argmin_{\phi, \theta} \mathcal{L}_{AE}^{\phi, \theta}(\mathbf{x}, \mathbf{\hat{x}}) = \mathcal{L}_{Rec}^{\phi, \theta}(\mathbf{x}, \mathbf{\hat{x}}) = \ell_1(\mathbf{x}, \mathbf{\hat{x}})
\end{equation}

, which in our case is the $\ell_1$-distance between input and reconstruction. The rationale behind this is the assumption that an AE trained on only healthy samples cannot properly reconstruct anomalies in pathological data. This approach has been successfully applied to anomaly segmentation in brain MRI~\cite{baur2018deep}\cite{atlason2019unsupervised} and in head CT~\cite{sato2018primitive}. A slightly different attempt was made in \cite{zimmerer2018context}, where the reconstruction-problem  was turned into an inpainting-task using a Context Autoencoder (Context AE) (Fig. \ref{fig:contextae}), in which the model is trained to recover missing sections in healthy training images. The natural choice for the shape of $\mathbf{z}$, here also referred to as \textit{latent space}, \textit{bottleneck} or \textit{manifold}, is a 1D vector. However, it has been shown that spatial AEs with a tensor-shaped bottleneck can be beneficial for high-resolution brain MRI as they preserve spatial context and can generate higher quality reconstructions~\cite{baur2018deep}.
%Please note that apart from shape, there is no regularization on the manifolds structure.

\textbf{Latent Variable Models}|In classic AEs, there is no regularization on the manifolds structure. In contrast, latent variable models such as Variational Autoencoders (VAEs\cite{Kingma2013AutoEncodingVB}, Fig. \ref{fig:vae}) constrain the latent space by leveraging the encoder and decoder networks of AEs to parameterize a latent distribution $q(\mathbf{z}) \sim \mathcal{N}(\mathbf{z}_{\mu}, \mathbf{z}_{\sigma})$, using the following objective:

\begin{align*}
    \argmin_{\phi, \theta} \mathcal{L}_{VAE}^{\phi, \theta}(\mathbf{x}, \mathbf{\hat{x}}) &= \mathcal{L}_{Rec}^{\phi, \theta}(\mathbf{x}, \mathbf{\hat{x}}) + \lambda_{KL}\mathcal{L}_{KL}^{\theta}(q(\mathbf{z}), p(\mathbf{z})) \\
		&= \ell_1(\mathbf{x}, \mathbf{\hat{x}}) + \lambda_{KL}\mathcal{D}_{KL}(q(\mathbf{z}) || p(\mathbf{z}))
\end{align*}

, where $\lambda_{KL}$ is a Lagrangian multiplier which weights the reconstruction loss against the distribution-matching KL-Divergence $\mathcal{D}_{KL}(\cdot || \cdot)$. In practice, the VAE projects input data onto a learned mean $\mathbf{\mu}$ and variance $\mathbf{\sigma}$, from which a sample is drawn and then reconstructed (see Fig. \ref{fig:vae}). While the VAE tries to match $q(\mathbf{z})$ to a prior $p(\mathbf{z})$ (typically a multivariate normal distribution) by minimizing the KL-Divergence, which has various shortcomings, the so-called Adversarial Autoencoder (AAE~\cite{44904}, Fig. \ref{fig:aae}) leverages an adversarial network as a proxy metric to minimize this discrepancy between the learned distribution $q(\mathbf{z})$ and the prior $p(\mathbf{z})$. As opposed to the KL-Divergence, the optimization via an adversarial network does not favor modes of distributions and is always differentiable. Another extension to the VAE, the so-called Gaussian Mixture VAE (GMVAE~\cite{dilokthanakul2016deep}) even replaces the mono-modal prior of the VAE with a gaussian mixture, leading to higher expressive power. Due to their ability to model the underlying distribution of high dimensional data, these frameworks are naturally suited for modeling the desired normative distribution. Further, their probabilistic nature facilitates the development of principled density-based anomaly detection methods. Consequently, they have been widely employed for outlier-based anomaly detection: VAEs were used in brain MRI for MS lesion~\cite{baur2018deep}, tumor and stroke detection~\cite{zimmerer2018context}. They have also been utilized for tumor detection in head CT~\cite{pawlowski2018unsupervised} from aggregate means of Monte-Carlo reconstructions. In brain MRI, AAE-\cite{chen2018unsupervised} and GMVAE\cite{you2019uad}-based approaches have also been successfully employed for tumor detection.

%To achieve this, the networks are trained to maximize the so-called evidence lower bound (ELBO), an approximation to the log-likelihood
%Naturally, their capability to model the underlying distribution of high dimensional data has been used to model healthy brain anatomy for detecting anomalies.

%all give rise to a principled probabilistic framework for modeling distributions with AEs by regularizing the latent space with a prior distribution. The models project input data onto a learned mean and variance, from which a sample is drawn and then reconstructed. To achieve this, $\mathbf{z}$ is regularized to follow a prior distribution, normally a Multivariate Normal Distribution (MVN).

%\begin{equation}
%    \mathcal{L}_{VAE} = -\mathbb{E}_{q(\mathbf{z}|\mathbf{x})} \left[ log \frac{p(\mathbf{x}|\mathbf{z})p(\mathbf{z})}{q(\mathbf{z}|\mathbf{x})} \right] = \mathcal{L}_{rec} + \mathcal{L}_{prior}    
%\end{equation}

%\begin{equation}
%    \mathcal{D}_{KL}(q(\mathbf{z}|\mathbf{x}) || p(\mathbf{z}))   
%\end{equation}

%\begin{equation}
%    \mathcal{L}_{rec} = -\mathbb{E}_{q(\mathbf{z}|\mathbf{x})}[log(p(\mathbf{x}|\mathbf{z}))]
%\end{equation}

\textbf{Generative Adversarial Networks}|Pioneering work, even before AEs were successfully applied for UAD in medical imaging, leveraged Generative Adversarial Networks (GANs~\cite{NIPS2014_5423}, Fig. \ref{fig:gan}) to detect anomalies in OCT data. Therefor, Schlegl et al ~\cite{schlegl2017unsupervised} modeled the distribution of healthy retinal patches with GANs and determined anomalies by computing the discrepancy between the retinal patch and a healthy counterpart restored by the GAN. Inspired by this work, Baur et al.~\cite{baur2018deep} leveraged the VAEGAN~\cite{larsen2015autoencoding}---a combination of the GAN and VAE (Fig. \ref{fig:anovaegan})---to overcome the training instabilities of the GAN and to allow for faster feed-forward inference, which they successfully employed for anomaly segmentation in brain MRI. In recent follow-up work, Schlegl et al.~\cite{schlegl2019f} improved on their GAN and also introduced an efficient way to replace the costly iterative restoration method by a single forward pass through the network.

%\min_{G} \max_{D} V(D,G) = \mathbb{E}_{x \sim p_{data}(x)}[log(D(x))] + \mathbb{E}_{z \sim p_z(z)}[1-log(D(G(z)))]

%\mathcal{L}_{VAEGAN} = \mathcal{L}_{rec} + \mathcal{L}_{prior} + \mathcal{L}_{GAN}
%\mathcal{L}_{rec} = -\mathbb{E}_{q(\mathbf{z}|\mathbf{x})}[log(p(\mathbf{x}|\mathbf{z}))]
%p(Dis_l(\mathbf{x})|\mathbf{z}) = \mathcal{N}(Dis_l(\mathbf{x})|Dis_l(\mathbf{\hat{x}}), \mathbf{I})

%\IEEEpubidadjcol

\subsection{Anomaly Segmentation}

The trained models can be used for anomaly detection \& segmentation in a variety of ways, which are summarized in the following. The interested reader is referred to the original papers for more detailed information.

\textbf{Reconstruction Based Methods}|Such approaches rely on pixel-wise residuals obtained from the difference

\begin{equation}
    \mathbf{r} = |\mathbf{x} - \mathbf{\hat{x}}|
    \label{eq:resmaps}
\end{equation}

of input samples $\mathbf{x}$ and their reconstruction $\mathbf{\hat{x}}$ (see Fig. \ref{fig:concept}). The underlying idea being that anomalous structures, which have never been seen during training, cannot be properly reconstructed from the distribution encoded in the latent space, such that reconstruction errors will be high for anomalous structures.

%-------------------

\begin{figure*}[!t]
\centering
\subfloat[Bayesian AE]{\includegraphics[width=2.5in]{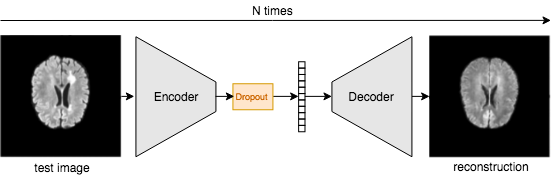}%
\label{fig:bae}}
\hfil
\subfloat[Bayesian VAE]{\includegraphics[width=2.5in]{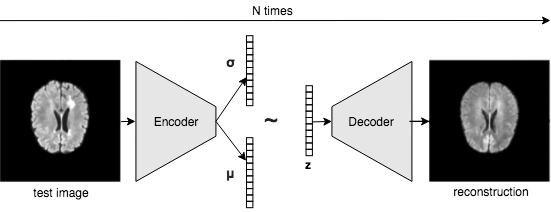}%
\label{fig:bvae}}
\caption{Monte Carlo Reconstructions aggregate and average N reconstructions for a single sample.}
\label{fig:mc_models}
\end{figure*}

\textbf{Monte Carlo Methods}|For non-deterministic generative models such as VAEs, multiple reconstructions can be obtained by Monte-Carlo (MC) sampling the latent space and an average consensus residual can be computed~\cite{pawlowski2018unsupervised}

\begin{equation}
    \mathbf{r} = \frac{1}{N} \sum_{n=1}^N |\mathbf{x} - \mathbf{\hat{x}_n}|
\end{equation}

, with $N$ being the number of MC samplings and $\mathbf{\hat{x}_n}$ being a single MC reconstruction.
For deterministic AEs, a similar effect can be achieved by applying dropout with rate $p_r$ to the latent space during inference time, which is also investigated in this work (see Fig. \ref{fig:bae} and Fig. \ref{fig:bvae} for a visual explanation).

\textbf{Gradient-Based Methods}|The gradient-based method proposed in \cite{zimmerer2018context} solely relies on image gradients obtained from a single backpropagation step when virtually optimizing for the following objective,

\begin{multline}
    \argmin_{\mathbf{\hat{x}}} \mathcal{L}_{Rec}(\mathbf{x}, \mathbf{\hat{x}}) + \lambda_{KL}\mathcal{L}_{KL}(\mathbf{z}, p(\mathbf{z})) \\
		= \ell_1(\mathbf{x}, \mathbf{\hat{x}}) + \lambda_{KL}\mathcal{D}_{KL}(\mathbf{z} || p(\mathbf{z}))
		\label{eq:grad}
\end{multline}

i.e. the pursuit of bringing the reconstruction $\mathbf{\hat{x}}$ and input $\mathbf{x}$ of the model together while simultaneously moving the latent representation of an input sample closer to the prior (the normal distribution). The resulting pixel-wise gradients are used as a saliency map for anomalies, where it is assumed that stronger gradients constitute anomalies.

\textbf{Restoration Based Methods}|In contrast to reconstruction based methods, restoration based methods involve an optimization on the latent manifold. In the pioneering approach using GANs \cite{schlegl2017unsupervised}, the goal is to iteratively move along the GANs input distribution $\mathbf{z}$ until a healthy variant of a query image is reconstructed well. Similarly, the method in \cite{you2019uad} tries to restore a healthy counterpart $\mathbf{\hat{x}}$ of an input sample $\mathbf{x}$, but by altering it until the ELBO of its latent representation $\mathbf{z}$ is maximized. This can be achieved by initializing $\mathbf{\hat{x}} = \mathbf{x}$ and then iteratively optimizing $\mathbf{\hat{x}}$ for the objective in Eq. \ref{eq:grad}. Again, the anomalies can be detected in image space from residual maps $\mathbf{r}$ (see Eq. \ref{eq:resmaps}).

\section{Experiments}
%Report really everything in a huge table and then extract and present pieces.

%Structure A
%-----------
%1. Detecting Hyper-Intense Lesions
%2. Detecting Space Occupying Lesions
%3. How much healthy training data is enough?
%4. Generalization / Domain Shift
%5. Original Architectures
%6. Uncertainty
%7. Clustering in Latent Space of Healthy and Anomalous Samples
%8. Meta-Discussion

%Structure B
%-----------
%1. Dense vs Spatial Bottleneck
%AE (dense) vs AE (spatial) vs GMVAE (dense) vs GMVAE (spatial)
%2. Constraining & Regularization
%AE (dense) vs VAE vs Constrained AE
% Introducing the constraint of similar latent representation of X and Xrec improved the performance.
%3. Monte-Carlo Methods
%AE (dense) vs Bayesian AE vs Bayesian VAE
%4. Latent Variable Models
%VAE vs GMVAE vs constrained AAE
% Using a GMM instead of a Unit Gaussian as Prior for the distribution of z did not improve the performance
%5. Reconstruction vs Restoration
% VAE vs VAE (restoration) vs GMVAE (spatial) vs GMVAE (spatial, restoration) vs f-AnoGAN
% Using an iterative restoration approach improved the performance.
%6. Domain Shift
%Performance of conceptually most different approaches on all datasets compared by AUPRC & Best possible dice
%-> or should we use all?
%-> AE (spatial), 
%7. Reconstruction Errors
% Compare all models (the non-original ones) in terms of their reconstruction error and try to correlate them to their actual UAD performance
%8. Model Complexity
%Compare the original methods the unified architecture in terms of their model complexity and their UAD performance

\begin{figure*}[!t]
\includegraphics[width=\textwidth]{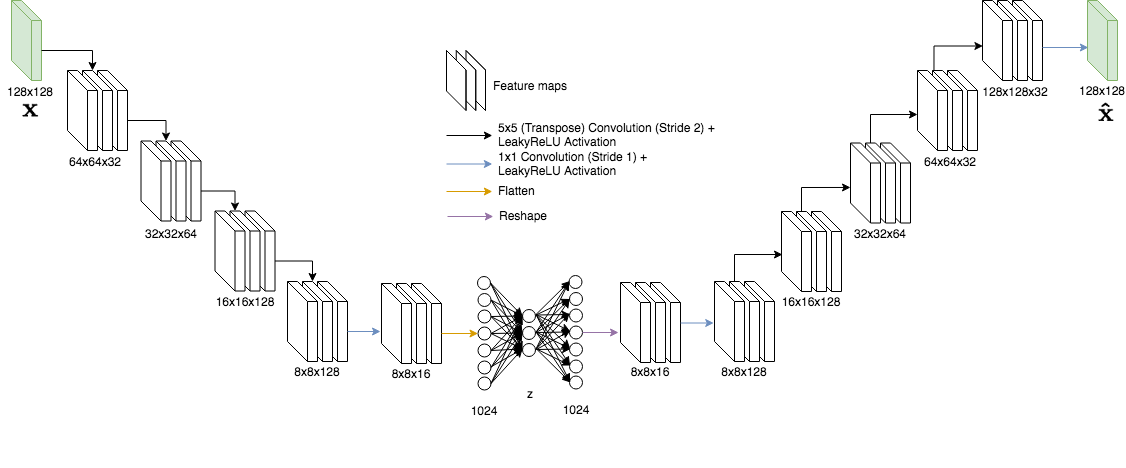}
\caption{The unified network architecture with a dense bottleneck. In the case of a spatial bottleneck, the flatten-, dense- and reshape-layers are replaced by a single set of 2D convolutional kernels.}
\label{fig:unified_architecture}
\end{figure*}

In the following, we first introduce the datasets used in the experiments, together with their pre-processing, and then introduce the unified network architecture which is the foundation of all the subsequently investigated models. We further explain our post-processing pipeline and all the metrics used in our investigations, before we finally present and discuss the results from various perspectives.

\subsection{Datasets}

\begin{table}[!t]
\renewcommand{\arraystretch}{1.3}
% if using array.sty, it might be a good idea to tweak the value of
% \extrarowheight as needed to properly center the text within the cells
\caption{Training, Validation, Testing subjects of the datasets used in this study}
\label{tab:datasets}
\centering
%% Some packages, such as MDW tools, offer better commands for making tables
%% than the plain LaTeX2e tabular which is used here.
\begin{tabular}{|l||c|c|c|}
\hline
\bfseries Dataset             & \bfseries Training   & \bfseries Validation & \bfseries Testing  \\ \hline
  $\mathcal{D}_{healthy}$       & 110                & 28             & -               \\ \hline
  $\mathcal{D}_{MS}$            & -                 & 3             & 45              \\ \hline
  $\mathcal{D}_{GB}$            & -                & -             & 28              \\ \hline
  $\mathcal{D}_{MSSEG2015}$     & -                 & -             & 20              \\ \hline
  %$\mathcal{D}_{MSSEG2008-CHB}$ & -                 & 2             & 8               \\
  %$\mathcal{D}_{MSSEG2008-UNC}$ & -                 & 2             & 8               \\
  $\mathcal{D}_{MSLUB}$         & -                 & -             & 30
  \\ \hline
\end{tabular}
\end{table}

For this survey, we rely on three different datasets. Selection criteria for these datasets were i) the availability of corresponding T1, T2 and FLAIR scans per subject to be able to leverage a single shared preprocessing pipeline and ii) each dataset being produced with a different MR device.

\textbf{Healthy, MS \& GB}|The primary dataset used in this comparative study is a homogenous set of MR scans of both healthy and diseased subjects, produced with a single Philips Achieva 3T MR scanner. It comprises FLAIR, T2- and T1-weighted MR scans of 138 healthy subjects, 48 subjects with MS lesions and 26 subjects with Glioma. All scans have been carefully reviewed and annotated by expert Neuro-Radiologists. Informed consent was waived by the local IRB.

%Patientwise split
%Training: 110 Subjects
%validation: 30 Subjects
%Testing: 48 Subjects (40 Testing, 8 Validation to choose an OP)

\textbf{MSLUB}|The second MRI dataset~\cite{lesjak2018novel} consists of co-registered T1, T2 and FLAIR scans of 30 different subjects with MS. Images have been acquired with a 3T Siemens Magnetom Trio MR system at the University Medical Center Ljubljana (UMCL). A gold standard segmentation was obtained from consensus segmentations of three expert raters.

\textbf{MSSEG2015}|The third MRI dataset in our experiments is the publicly available training set of the 2015 Longitudinal MS lesion segmentation challenge~\cite{carass2017longitudinal}, which contains 21 scan sessions from 5 different subjects with T1, T2, PD and FLAIR images each. All data has been acquired with a 3.0 Tesla Philips MRI scanner. The exact device is not known, but the intensity distribution is different from our primary MS \& GB datasets. Thus, in this study we utilize the data to test the generalization capabilities of the models and approaches.

% What about BRATS? SHould we do it?

%TODO: REFORMULATE
%MSSEG2008:  The data acquired at University of North Carolina ($\mathcal{D}_{MSSEG2008-UNC}$) and the Children's Hospital Boston ($\mathcal{D}_{MSSEG2008-CHB}$) comprises FLAIR, T1 and T2-weighted images from 10 subjects per site.
% TODO END

\textbf{Preprocessing and Split}|All scans have been brought to the SRI24 ATLAS~\cite{Rohlfing:2009dp} space to ensure all data share the same volume size and orientation. In succession, the scans have been skull-stripped with ROBEX~\cite{Iglesias:2011fb} and denoised with CurvatureFlow~\cite{sethian1999level}. Prior to feeding the data to the networks, all volumes have been normalized into the range [0,1] by dividing each scan by its 98th percentile. All datasets have randomly been split (patient-wise) into training, validation and testing sets as listed in Table \ref{tab:datasets}. Training and testing is done on all axial slices of each volume for which the corresponding brainmask indicates the presence of brain pixels. Modus operandi is at a slice resolution of $128\times 128$px. This is in stark contrast to some other works, which restrict themselves anatomically to the axial midline~\cite{baur2018deep} or lower resolution~\cite{pawlowski2018unsupervised,chen2018unsupervised}.
\subsection{Network Architecture and Models}

The unified architecture depicted in Fig. \ref{fig:unified_architecture} was empirically determined in a manual iterative architecture search. The goal was to achieve low reconstruction error on both the training and validation data from $\mathcal{D}_{healthy}$. This unified architecture was then used to train a great variety of models coming from the different, previously introduced domains:

\textbf{Autoencoders}|As a baseline, we used the unified architecture to train a variety of non-generative AEs:

\begin{enumerate}
    \item \textbf{AE (dense)}: an AE with a dense bottleneck $\mathbf{z} \in \mathcal{R}^{128}$
    \item \textbf{AE (spatial)}~\cite{baur2018deep}: an AE with a spatial bottleneck $\mathbf{z} \in \mathcal{R}^{8\times 8 \times 128}$
    \item \textbf{Context AE}~\cite{zimmerer2018context}: with $\mathbf{z} \in \mathcal{R}^{128}$
    \item \textbf{Constrained AE}~\cite{chen2018unsupervised}: with $\mathbf{z} \in \mathcal{R}^{128}$
\end{enumerate}

\textbf{Latent Variable Models}|Further, we trained various generative latent variable models using the same unified architecture and bottleneck configurations:

\begin{enumerate}
    \item \textbf{VAE}~\cite{baur2018deep,zimmerer2019unsupervised}: with $\mathbf{z} \in \mathcal{R}^{128}$
    \item \textbf{Context VAE}~\cite{zimmerer2018context}: with $\mathbf{z} \in \mathcal{R}^{128}$
    \item \textbf{Constrained AAE}~\cite{chen2018unsupervised}: with $\mathbf{z} \in \mathcal{R}^{128}$
    \item \textbf{GMVAE (dense)}~\cite{you2019uad}: with $\mathbf{z} \in \mathcal{R}^{128}$
    \item \textbf{GMVAE (spatial)}~\cite{you2019uad}: with $\mathbf{z} \in \mathcal{R}^{8\times 8 \times 128}$
\end{enumerate}

%an AE with a dense bottleneck $\mathbf{z} \in \mathcal{R}^{128}$ (refererred to as \textit{AE (dense)}), with a spatial bottleneck $\mathbf{z} \in \mathcal{R}^{8\times 8 \times 128}$ (referred to as \textit{AE (spatial)}, and both a Context AE and Constrained AE with the same dense bottleneck.
%, a dense VAE, both a dense and spatial GMVAE (GMVAE$_{d}$ and AE$_{s}$, respectively), an fAnoGAN, a CAE, a CVAE and a CAAE. 

\textbf{Generative Adversarial Networks}|Finally, we also trained an AnoVAEGAN~\cite{baur2018deep} and an f-AnoGAN~\cite{schlegl2019f}, whose encoder-decoder networks implement the unified architecture, and the discriminator network is a replica of the encoder:

\begin{enumerate}
    \item \textbf{AnoVAEGAN}~\cite{baur2018deep}: with $\mathbf{z} \in \mathcal{R}^{128}$
    \item \textbf{fAnoGAN}~\cite{schlegl2019f}: with $\mathbf{z} \in \mathcal{R}^{128}$
\end{enumerate}

Noteworthy, both methods were optimized with the Wasserstein loss~\cite{pmlr-v70-arjovsky17a} to avoid GAN training instabilities and mode collapse.

All models were trained from $\mathcal{D}_{healthy}$ until convergence using an automatic early stopping criterion, i.e. training was stopped if the reconstruction loss on the held-out validation set from $\mathcal{D}_{healthy}$ did not improve more than an $\epsilon > 10e-9$ for 5 epochs. In succession, all the methods were used for reconstruction-based anomaly detection. The trained VAE and GMVAE were also used for the density-based image restoration~\cite{you2019uad}, where each sample was restored in 500 iterations:

\begin{enumerate}
    \item \textbf{VAE (restoration)}~\cite{you2019uad}
    \item \textbf{GMVAE (restoration)}~\cite{you2019uad}
\end{enumerate}

Both AE (dense) and VAE were also used for MC-reconstruction based anomaly detection:

\begin{enumerate}
    \item \textbf{Bayesian AE}~\cite{pawlowski2018unsupervised}: Dropout rate 0.2
    \item \textbf{Bayesian VAE}~\cite{pawlowski2018unsupervised}: $N=100$ MC-samples per input slice
\end{enumerate}

and in the case of the Context VAE, we also tried the gradient-based approach proposed in \cite{zimmerer2018context}:

\begin{enumerate}
    \item \textbf{Context VAE (gradient)}~\cite{zimmerer2018context}
\end{enumerate}

Hyperparameters can be taken from Table \ref{tab:hyperparams}.

\begin{table}[!t]
\renewcommand{\arraystretch}{1.3}
% if using array.sty, it might be a good idea to tweak the value of
% \extrarowheight as needed to properly center the text within the cells
\caption{Hyperparameters for the different models}
\label{tab:hyperparams}
\centering
%% Some packages, such as MDW tools, offer better commands for making tables
%% than the plain LaTeX2e tabular which is used here.
\begin{tabular}{|l||c|}
\hline
\bfseries Param             & \bfseries Value   \\ \hline
  learning rate       & 0.0001  \\ \hline
  $\lambda_{KL}$      & 1.0 										\\ \hline
	dropout rate $p_r$      & 0.2
  \\ \hline
\end{tabular}
\end{table}
\subsection{Postprocessing}

The output of all models and approaches is subject to the same post-processing. Every residual image $\mathbf{r}$ is first multiplied with a slightly eroded brain-mask to remove prominent residuals occuring near sharp edges at brain-mask boundaries and gyri and sulci (the latter are very diverse and hard to model). Further, for the MS lesion datasets we make use of prior knowledge and only keep positive residuals as these lesions are known to be fully hyper-intense in FLAIR images. For each MR volume, the residual images for all slices are first aggregated into a corresponding 3D residual volume, which is then subject to a 3D median filtering with a $5\times 5\times 5$ kernel to remove small outliers and to obtain a more continuous signal. The latter is beneficial for the subsequent model assessment as it leads to smoother curves. As a final step, the continuous output is binarized and a 3D connected component analysis is performed on the resulting binary volumes to discard any small structures with an area less than 8 voxels.
\subsection{Metrics}

We assess the anomaly segmentation performance at a level of single voxels, at which class imbalance needs careful consideration as anomalous voxels are usually less frequent than normal voxels. To do so, we generate dataset-specific Precision-Recall-Curves (PRC) and then compute the area under it (AUPRC). Noteworthy, this allows to judge the models capabilities without choosing an Operating Point (OP). Further, for each model we provide an estimate of its theoretically best possible DICE-score ($\lceil$DICE$\rceil$) on each dataset. Therefor, for each testing dataset $d \in \mathcal{D} = {\mathcal{D}_{MS}, \mathcal{D}_{GB}, \mathcal{D}_{MSSEG2015}, \mathcal{D}_{MSLUB}}$, we utilize the available ground-truth segmentation and perform a greedy search up to three decimals to determine the respective OP on the PRC curve which yields the best possible DICE score for dataset $d$. Additionally, to simulate the models performance in more realistic settings, we utilize a held-out validation set from $\mathcal{D}_{MS}$ to determine an OP $t$ at which we then compute patient-specific DICE-scores for every dataset. Some of the reviewed works originally utilize Receiver-Operating-Characteristics (ROC) to evaluate anomaly detection performance. We report the area under such ROC curves (AUROC) as well, but want to emphasize that it has to be used with care. Under heavy class imbalance, ROC curves can be misleading as they give much higher weight to the more frequent class and thus in the case of a pixel-wise assessment very optimistic views on performance. 

To gain deeper insights what makes a model capable of segmenting anomalies better than others, we also report dataset-specific $\ell_1$-reconstruction errors on normal ($\ell_1$-RE$_N$) and anomalous voxels ($\ell_1$-RE$_A$), as well as the $\mathcal{X}^2$-distance of the respective normal and anomalous residual histograms for every model.

%For a relative comparison between these reconstruction errors, we also compute the intersection of the histogram $H_A$ of anomalous pixels and the histogram $H_N$ of normal intensities (inside the brainmask, both normalized by the total number of pixels of both histograms combined), extracted from the residual images $\mathbf{r}$

%\begin{equation}
%    HIS_r = \sum_i \min(H_A^i, H_N^i)
%\end{equation}

%, where $H_A^i$ and $H_N^i$ refer to the i-th bin, respectively.

% https://stats.stackexchange.com/questions/7400/how-to-assess-the-similarity-of-two-histograms
% https://stats.stackexchange.com/questions/123445/evaluate-overlap-of-two-histograms-which-are-normalized-to-area-1
\subsection{Overview}

% See https://texblog.org/2012/05/30/generate-latex-tables-from-csv-files-excel/
%\csvautotabular{csv/all_results.csv}
\begin{table*}[!t]
\caption{Experimental results on the MS dataset}
\label{tab:dms_results}
\centering
\csvreader[tabular=|l|l|l|l|l|l|l|l|,
    table head=\hline \textbf{Approach} & \textbf{AUROC} & \textbf{AUPRC} & $\lceil$\textbf{DICE}$\rceil$ & \textbf{DICE ($\mu \pm \sigma$)} & \textbf{$\ell_1$-RE$_N$ ($\mu \pm \sigma$)} & \textbf{$\ell_1$-RE$_A$ ($\mu \pm \sigma$)} & $\mathcal{X}^2$ \\ \hline,
    late after line=\\\hline]%
{csv/all_results_unified.csv}{Approach=\approach, MSKRI AUROC=\aurocms, MSKRI AUPRC=\auprcms, MSKRI BPDICE=\bpdicems, MSKRI DICE=\dicems, MSKRI Rec.-Error (Normal)=\recerrnms, MSKRI Rec.-Error (Anomalous)=\recerrams, MSKRI IOU=\ioums, MSKRI ChiSq=\chisqms}%
{\approach & \aurocms & \auprcms & \bpdicems & \dicems & \recerrnms & \recerrams & \chisqms}%
\end{table*}

\begin{table*}[!t]
\caption{Experimental results on the GB dataset}
\label{tab:dgb_results}
\centering
\csvreader[tabular=|l|l|l|l|l|l|l|l|,
    table head=\hline \textbf{Approach} & \textbf{AUROC} & \textbf{AUPRC} & $\lceil$\textbf{DICE}$\rceil$ & \textbf{DICE ($\mu \pm \sigma$)} & \textbf{$\ell_1$-RE$_N$ ($\mu \pm \sigma$)} & \textbf{$\ell_1$-RE$_A$ ($\mu \pm \sigma$)} & $\mathcal{X}^2$ \\ \hline,
    late after line=\\\hline]%
{csv/all_results_unified.csv}{Approach=\approach, GBKRI AUROC=\aurocgb, GBKRI AUPRC=\auprcgb, GBKRI BPDICE=\bpdicegb, GBKRI DICE=\dicegb, GBKRI Rec.-Error (Normal)=\recerrngb, GBKRI Rec.-Error (Anomalous)=\recerragb, GBKRI IOU=\iougb, GBKRI ChiSq=\chisqgb}%
{\approach & \aurocgb & \auprcgb & \bpdicegb & \dicegb & \recerrngb & \recerragb & \chisqgb}%
\end{table*}

\begin{table*}[!t]
\caption{Experimental results on the MSLUB dataset}
\label{tab:dmslub_results}
\centering
\csvreader[tabular=|l|l|l|l|l|l|l|l|,
    table head=\hline \textbf{Approach} & \textbf{AUROC} & \textbf{AUPRC} & $\lceil$\textbf{DICE}$\rceil$ & \textbf{DICE ($\mu \pm \sigma$)} & \textbf{$\ell_1$-RE$_N$ ($\mu \pm \sigma$)} & \textbf{$\ell_1$-RE$_A$ ($\mu \pm \sigma$)} & $\mathcal{X}^2$ \\ \hline,
    late after line=\\\hline]%
{csv/all_results_unified.csv}{Approach=\approach, MSLUB AUROC=\aurocmslub, MSLUB AUPRC=\auprcmslub, MSLUB BPDICE=\bpdicemslub, MSLUB DICE=\dicemslub, MSLUB Rec.-Error (Normal)=\recerrnmslub, MSLUB Rec.-Error (Anomalous)=\recerramslub, MSLUB IOU=\ioumslub, MSLUB ChiSq=\chisqmslub}%
{\approach & \aurocmslub & \auprcmslub & \bpdicemslub & \dicemslub & \recerrnmslub & \recerramslub & \chisqmslub}%
\end{table*}

\begin{table*}[!t]
\caption{Experimental results on the MSSEG2015 dataset}
\label{tab:dmsseg2015_results}
\centering
\csvreader[tabular=|l|l|l|l|l|l|l|l|,
    table head=\hline \textbf{Approach} & \textbf{AUROC} & \textbf{AUPRC} & $\lceil$\textbf{DICE}$\rceil$ & \textbf{DICE ($\mu \pm \sigma$)} & \textbf{$\ell_1$-RE$_N$ ($\mu \pm \sigma$)} & \textbf{$\ell_1$-RE$_A$ ($\mu \pm \sigma$)} & $\mathcal{X}^2$ \\ \hline,
    late after line=\\\hline]%
{csv/all_results_unified.csv}{Approach=\approach, ISBI AUROC=\aurocmsseg2015, ISBI AUPRC=\auprcmsseg2015, ISBI BPDICE=\bpdicemsseg2015, ISBI DICE=\dicemsseg2015, ISBI Rec.-Error (Normal)=\recerrnmsseg2015, ISBI Rec.-Error (Anomalous)=\recerramsseg2015, ISBI IOU=\ioumsseg2015, ISBI ChiSq=\chisqmsseg2015}%
{\approach & \aurocmsseg2015 & \auprcmsseg2015 & \bpdicemsseg2015 & \dicemsseg2015 & \recerrnmsseg2015 & \recerramsseg2015  & \chisqmsseg2015}%
\end{table*}

\begin{figure*}[!t]
\centering
\subfloat[MSKRI]{\includegraphics[width=0.50\textwidth]{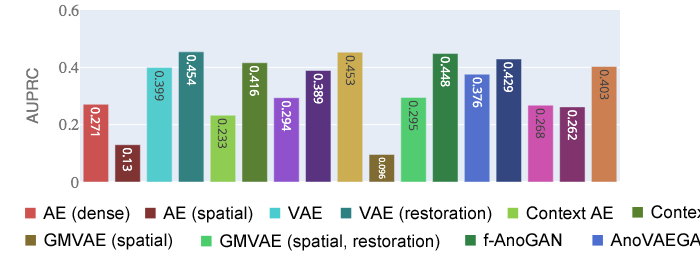}%
\label{fig:barchart_auprc_unified_mskri}}
\hfil
\subfloat[GBKRI]{\includegraphics[width=0.50\textwidth]{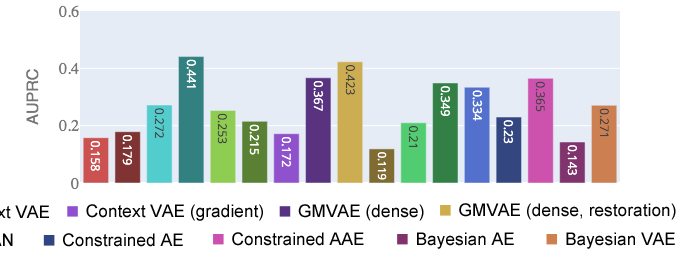}%
\label{fig:barchart_auprc_unified_gbkri}}

\subfloat[MSLUB]{\includegraphics[width=0.50\textwidth]{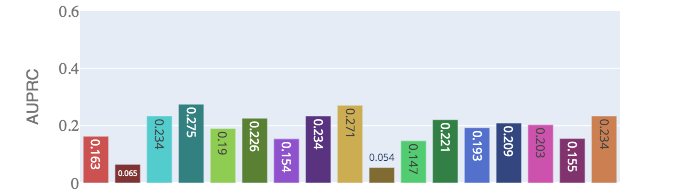}%
\label{fig:barchart_auprc_unified_mslub}}
\hfil
\subfloat[MSSEG2015]{\includegraphics[width=0.50\textwidth]{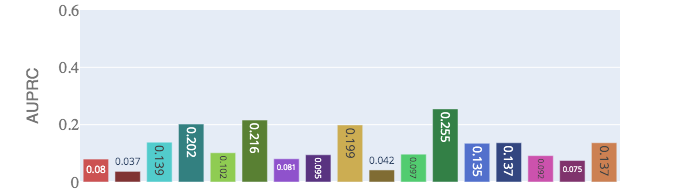}%
\label{fig:barchart_auprc_unified_msseg2015}}

\caption{AUPRC of all models and UAD approaches, using the unified architecture.}
\label{fig:unified_barchart}
\end{figure*}

%Figure: All models (unified architecture) next to each other. The non-unified variants are plotted separately

Detailed results of all models and UAD approaches on all datasets can be found in Tables \ref{tab:dms_results} ($\mathcal{D}_{MS}$), \ref{tab:dgb_results} ($\mathcal{D}_{GB}$), \ref{tab:dmslub_results} ($\mathcal{D}_{MSLUB}$) and \ref{tab:dmsseg2015_results} ($\mathcal{D}_{MSSEG2015}$). In the following, we analyze all these data from different perspectives. We start by first comparing different model types and bottleneck design, followed by the different ways to detect anomalies directly in image-space. Then, we shed the light on the number of training subjects and their impact on performance, and elaborate on domain shift.

%domain shift, number of training subjects and their impact on performance, model types and different UAD approaches.

\begin{figure*}[!t]
\centering
\subfloat[axial slice from $\mathcal{D}_{MS}$, ventral]{\includegraphics[width=\textwidth]{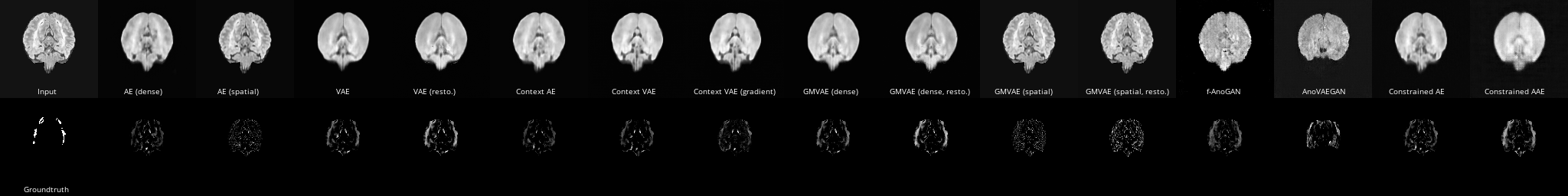}%
\label{fig:examples_mskri_ventral}}
\hfil
\subfloat[axial slice from $\mathcal{D}_{MS}$, midline]{\includegraphics[width=\textwidth]{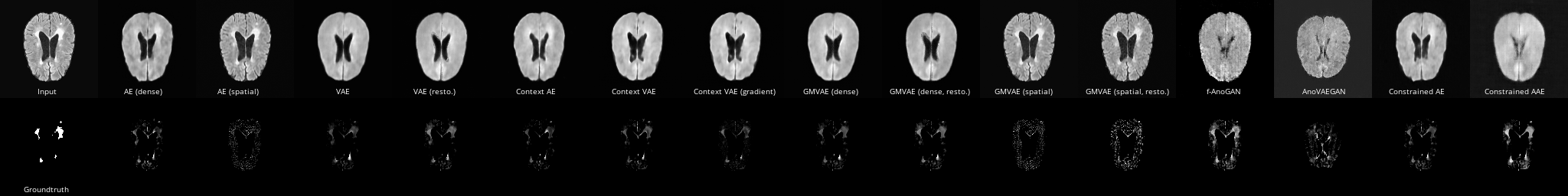}%
\label{fig:examples_mskri_midline}}
\hfil
\subfloat[axial slice from $\mathcal{D}_{MS}$, dorsal]{\includegraphics[width=\textwidth]{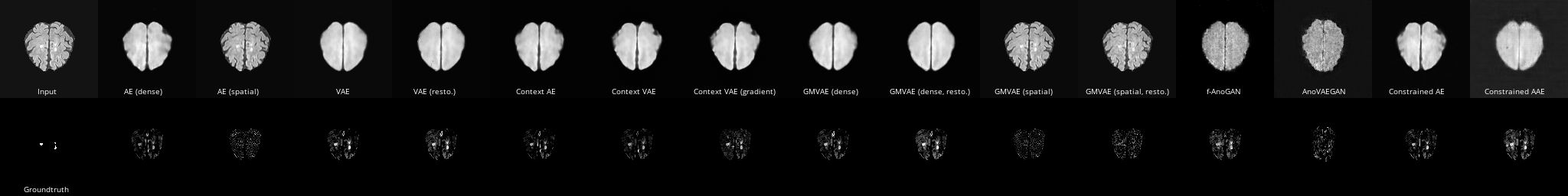}%
\label{fig:examples_mskri_dorsal}}
\hfil
\subfloat[axial slice from $\mathcal{D}_{GB}$, ventral]{\includegraphics[width=\textwidth]{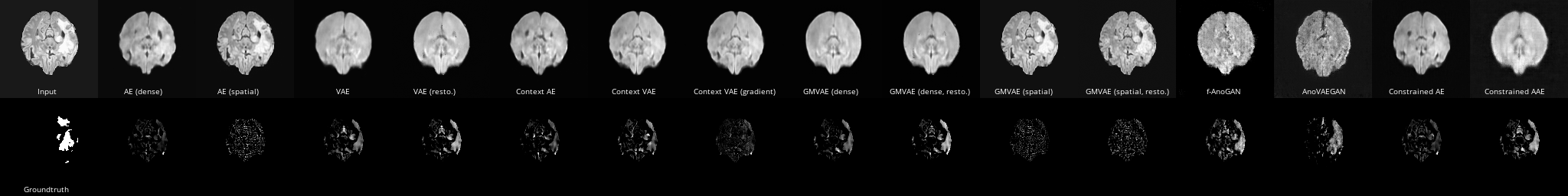}%
\label{fig:examples_gbkri_ventral}}
\hfil
\subfloat[axial slice from $\mathcal{D}_{MSSEG2015}$, dorsal]{\includegraphics[width=\textwidth]{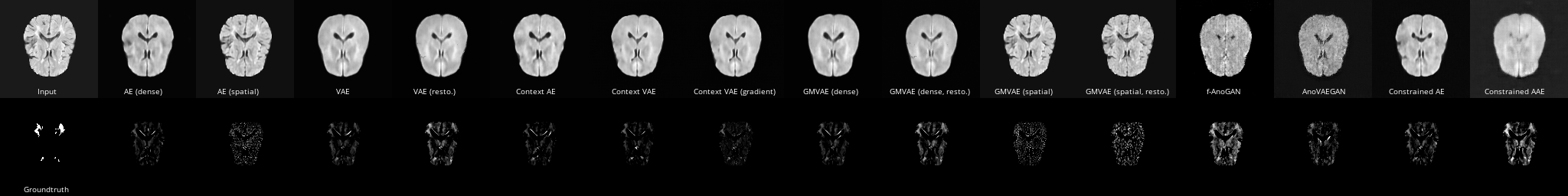}%
\label{fig:examples_isbi_ventral}}
\hfil
\subfloat[axial slice from $\mathcal{D}_{MSLUB}$, midline]{\includegraphics[width=\textwidth]{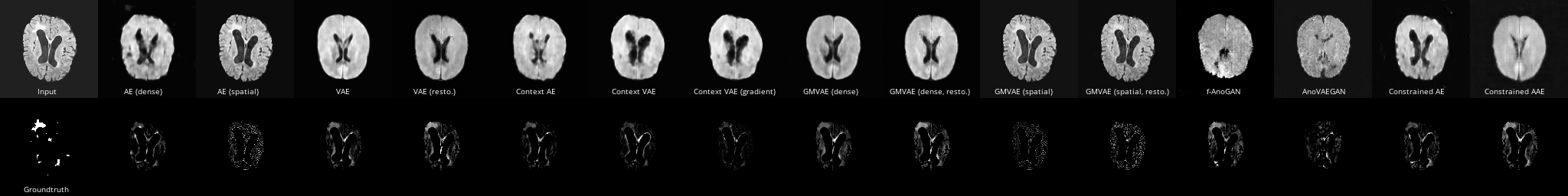}%
\label{fig:examples_mslub_midline}}

\caption{Visual examples of the different reviewed methods on different datasets, using the unified architecture. Top row: reconstructions; Bottom row: raw residuals.}
\label{fig:examples_unified}
\end{figure*}

\subsection{Constraining \& Regularization}

% Figure which hows the AUPRC, $\lceil$DICE$\rceil$, Rec-Err (Normal) and Rec-Err (Anomalous) for the 3 models on MS lesions

\begin{figure*}[!h]
\centering
\includegraphics[width=\textwidth]{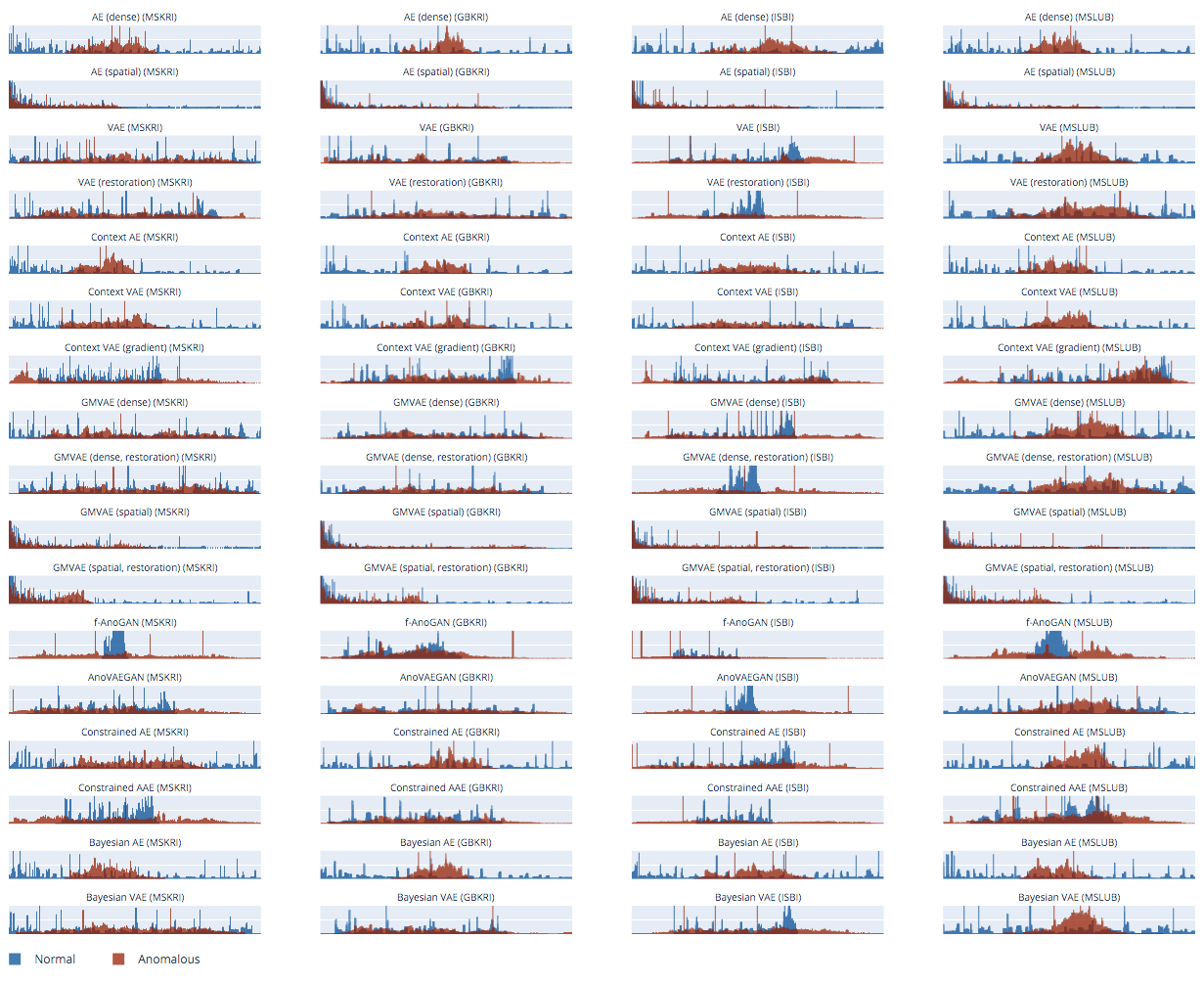}%
\caption{Normalized histograms of residuals of normal (blue) and anomalous (red) pixels in the intensity range $\in ]0;1.0]$ (ignoring residuals which are completely 0)}
\label{fig:hist}
\end{figure*}

Initially, we compare the classic AE (dense) to its VAE and Constrained AE counterpart to investigate the effect of constraining or regularizing the latent space of the models. Recall that VAEs regularize the latent space to follow a prior distribution, whereas the deterministic Constrained AE enforces that reconstructions and input lie closely on the manifold. We measure the models' performances in terms of the AUPRC as well as the $\lceil$DICE$\rceil$ and glimpse at the reconstruction errors for normal and anomalous pixels (see Fig. \ref{fig:hist} for residual histograms of normal and anomalous voxels). We see from Table \ref{tab:dms_results} that explicitly modeling a distribution with a VAE leads to dramatic performance gains on $\mathcal{D}_{MS}$ over the standard AE, and introducing the matching constraint (Constrained AE) between $\mathbf{x}$ and $\hat{\mathbf{x}}$ improves the performance even more. On all other datasets, the VAE clearly is the winner among the compared models, but the Constrained AE still outperforms the classic AE. From these results, we deduce that enforcing a structure on the manifold of AEs is indeed beneficial for UAD.
%, but at the same time a principled variational modeling of a distribution (approximating a manually chosen prior) is not always mandatory.

\subsection{Dense vs Spatial Bottleneck}

% Figure which shows the AUPRC, $\lceil$DICE$\rceil$, Rec-Err (Normal) and Rec-Err (Anomalous) for the 4 models on MS lesions
\begin{figure}
\centering
\includegraphics[width=2.5in]{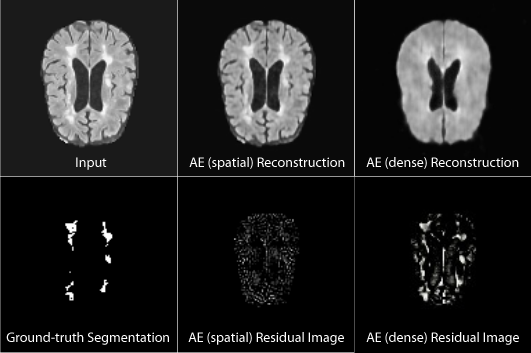}%
\caption{Reconstructions and postprocessed residuals using dense and spatial AEs}
\label{fig:dense_vs_spatial}
\end{figure}

To determine if the design of the AE bottleneck can improve the performance of the models, we further compare dense models for which a spatial counterpart exists, i.e. AE (dense) vs AE (spatial) vs GMVAE (dense) vs GMVAE (spatial). The spatial bottleneck allows the model to preserve spatial information and geometric features in its latent space, which positively affects the models reconstruction capabilities. From Tables \ref{tab:dms_results}-\ref{tab:dmsseg2015_results} it can be seen that the dense models outperform the spatial variants, alone the spatial AE performs slightly better on $\mathcal{D}_{GB}$ than its dense counterpart. We find that at our resolution of 128x128px, the spatial models reconstruct their input too well (see Fig. \ref{fig:dense_vs_spatial}), including the anomalies.

%(as opposed to the findings reported in \cite{baur2018deep} at a resolution of $256\times 256$px).

%To determine if the design of the AE bottleneck can boost the performance of the models, we further compare dense models for which a spatial counterpart exists, i.e. AE (dense) vs AE (spatial) vs GMVAE (dense) vs GMVAE (spatial), in terms of their AUPRC and $\lceil$DICE$\rceil$. From the tables \ref{tab:dms_results}-\ref{tab:dmsseg2015_results} we see that in all cases the dense models significantly outperform the spatial variants across all datasets, regardless of the model being deterministic or probabilistic. The reason is that at a resolution of 128x128px, the spatial models reconstruct the input too well, including the anomalies (see Fig. \ref{fig:dense_vs_spatial}). This is in contrast to the findings reported at a resolution of $256\times 256$px in \cite{baur2018deep}, where spatial methods seem to outperform the dense variants.

\subsection{Latent Variable Models}

Next, we focus only on different latent variable model types, i.e. the VAE, GMVAE (dense) and Constrained AAE. On the MS datasets $\mathcal{D}_{MS}, \mathcal{D}_{MSSEG2015}$ and $\mathcal{D}_{MSLUB}$, the VAE constitutes the best among the compared models. The Constrained AAE yields lower performance than the other models---also lower than its non-generative sibling, the Constrained AE. However, on the Glioblastoma dataset, it is on par with the GMVAE, and both models significantly outperform the VAE in the detection of brain tumors. Generally, the performance of the GMVAE generally seems to heavily depend on the dataset rather than the pathology: On 
$\mathcal{D}_{MS}$ and $\mathcal{D}_{MSLUB}$ it behaves very similar to the VAE, whereas on $\mathcal{D}_{GB}$ and $\mathcal{D}_{MSSEG2015}$ its performance resembles that of the Constrained AAE.

\subsection{GAN-based models}

GAN-based models are known to produce very realistic and crisp images, while AEs are known for their blurry reconstructions. Indeed, qualitative comparison of the f-AnoGAN and the AnoVAEGAN to the AE and VAE shows that the GAN-based models promote sharpness. This is particularly evident near the boundaries of the brain (see Fig. \ref{fig:examples_unified}). However, both the f-AnoGAN and AnoVAEGAN model the training distribution too well, such that reconstructions often differ anatomically from the actual input samples (see Fig. \ref{fig:examples_mskri_midline} for an axial midline slice from $\mathcal{D}_{MS}$). This is especially the case for the AnoVAEGAN, which produces the most crisp reconstructions, but often does not preserve anatomical coherence at all. As a result, on the MS datasets its performance is only comparable to the VAE, but it works considerably better for Glioblastoma segmentation. The f-AnoGAN does not provide as crisp images, but preserves the shape of the input sample and the difference between reconstruction residuals on normal and anomalous pixels is considerably higher across all datasets than for any of the other methods. This makes the UAD performance of the f-AnoGAN stand out. In total, both GAN-based approaches significantly outperform the standard AE (on average, more than 9\% for the AnoVAEGAN and more than 15\% for the f-AnoGAN) and the f-AnoGAN clearly also outperforms the VAE (on average more than 6\%).

\subsection{Monte-Carlo Methods}

% Figure which shows AE (dense) vs Bayesian AE and VAE vs Bayesian VAE with AUPRC, $\lceil$DICE$\rceil$, Rec-Err (Normal) and Rec-Err (Anomalous)
Monte-Carlo methods applied to (variational) AEs provide an interesting means to aggregate a consensus reconstruction, in which only very likely image features should be emphasized. To investigate if anomalies are affected, we experiment with $N=100$ MC-reconstructions and---where necessary---an empirically chosen dropout-rate $p_d = 0.2$ to trade-off reconstruction quality and chance. We find that, compared to one-shot reconstructions, the impact of MC-sampling is at most subtle, and not consistent across different models and datasets. A comparison of AE (dense) to the Bayesian AE shows that MC-dropout leads to a slightly worse performance in almost all metrics across all datasets. On the other hand, the Bayesian VAE, which does not need dropout for MC sampling due to its probabilistic bottleneck, is equal to or slightly outperforms the VAE on $\mathcal{D}_{MS}$, but not on $\mathcal{D}_{GB}$ and $\mathcal{D}_{MSLUB}$. Overall, these numbers indicate that MC methods, albeit an interesting approach, do not provide significant gains in the way they are currently employed. 

\subsection{Reconstruction vs Restoration}

Previous comparisons focused on different model types and all relied on the reconstruction-based UAD concept. In the following, we rank reconstruction-based methods, against gradient- and restoration-based UAD approaches. More precisely, we compare reconstruction against restoration on the VAE, GMVAE (dense) and GMVAE (spatial). We further rank the restoration-based methods against the top-candidate f-AnoGAN. From Tables \ref{tab:dms_results} to \ref{tab:dmsseg2015_results} it is evident that restoration based UAD is generally superior to the reconstruction-based counterparts (ranging from 4-17\% for the VAE, 4-10\% for the dense GMVAE and 6-20\% for the spatial GMVAE). Consistent with our previously measured results on dense versus spatial models, we also witness a dramatic drop in performance when using the spatial GMVAE, though. Except for $\mathcal{D}_{MSSEG2015}$, the dense restoration methods outperform the f-AnoGAN in all scenarios in terms of the AUPRC and $\lceil$DICE$\rceil$.

\subsection{Domain Shift}

Deep Learning models trained from data coming from one domain generally have difficulties to generalize well to other domains, and tackling such domain shift is still a highly active research area. Here, we want to determine to which extent AEs are prone to this effect and if some methods generalize better than others. Subject to our investigations are the MS datasets $\mathcal{D}_{MS}$, $\mathcal{D}_{MSLUB}$ and $\mathcal{D}_{MSSEG2015}$ among which such shifts occur. Generally, UAD performance is best on $\mathcal{D}_{MS}$, which matches the training data distribution, and on both $\mathcal{D}_{MSLUB}$ \& $\mathcal{D}_{MSSEG2015}$, the UAD performance drops significantly. However, the reasons for this drop can be manifold, and we want to emphasize that UAD performance as such is not a good indicator for domain shift, as the lesion size and count differs across datasets, and the contrast for $\mathcal{D}_{MS}$ is considerably better than for the other datasets. Instead, we suggest to look at the reconstruction error of normal pixels $\ell_1$-RE$_N$ in these datasets. From Tables \ref{tab:dms_results}, \ref{tab:dmslub_results} and \ref{tab:dmsseg2015_results} it can be seen that this error hardly degrades across all these datasets. This implies that generalization measured in terms of the models reconstruction capabilities is not of primary concern. However, from aforementioned tables it can be seen that the reconstruction error of anomalous pixels $\ell_1$-RE$_N$ is significantly smaller on $\mathcal{D}_{MSLUB}$ and $\mathcal{D}_{MSSEG2015}$, which is a clear indicator of weaker contrast between normal tissue and lesions in these datasets.

\subsection{Different Pathologies}

On both Multiple Sclerosis ($\mathcal{D}_{MS}$) and Glioblastoma ($\mathcal{D}_{GB}$), the restoration-based approaches with dense bottleneck constitute the top-performers, delivering results in roughly the same league. Similarly, lowest performances can be seen from the spatial models, the gradient-based UAD approach and the standard AE. However, in contrast to $\mathcal{D}_{MS}$, on $\mathcal{D}_{GB}$ there is a large performance gap between the top-performing restoration approaches and any other methods: the GAN-based methods f-AnoGAN and AnoVAEGAN drop by 10\% and 4\%, respectively, the performance of the VAE models degrades by at least 12\% and the Constrained AE even loses 20\% in AUPRC. Interestingly, the Constrained AAE gains by 10\%. Multiple factors lead to the lower performance: In contrast to MS lesions, tumors do not purely appear hyper-intense in FLAIR MRI. Some compartments of the tumor also resemble normal tissue, and the investigated UAD approaches have difficulties to properly delineate those. Second, tumors often are not only larger than MS lesions, but can have very complex shape (see Fig. \ref{fig:examples_gbkri_ventral}). This is hard to segment with precision---even among human annotators, there is variation.

\subsection{How much healthy training data is enough?}

\begin{figure*}[!t]
\centering
\subfloat[MSKRI]{\includegraphics[width=0.50\textwidth]{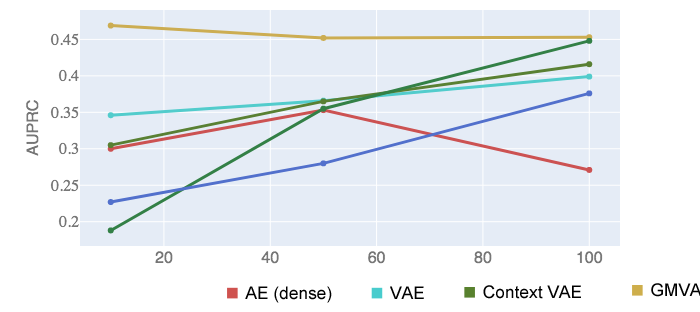}%
\label{fig:lineplot_auprc_numpatients_mskri}}
\hfil
\subfloat[GBKRI]{\includegraphics[width=0.50\textwidth]{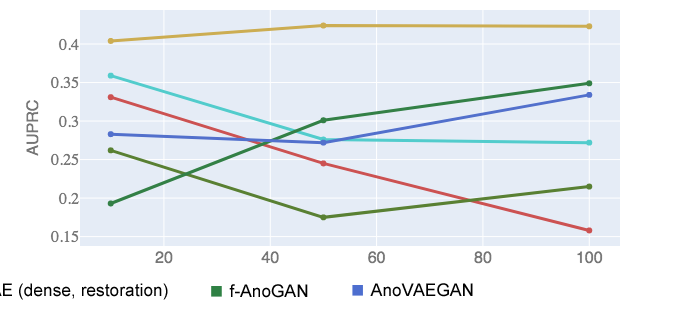}%
\label{fig:lineplot_auprc_numpatients_gbkri}}

\subfloat[MSLUB]{\includegraphics[width=0.50\textwidth]{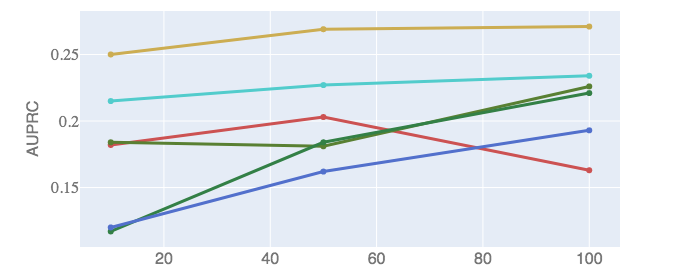}%
\label{fig:lineplot_auprc_numpatients_mslub}}
\hfil
\subfloat[MSSEG2015]{\includegraphics[width=0.50\textwidth]{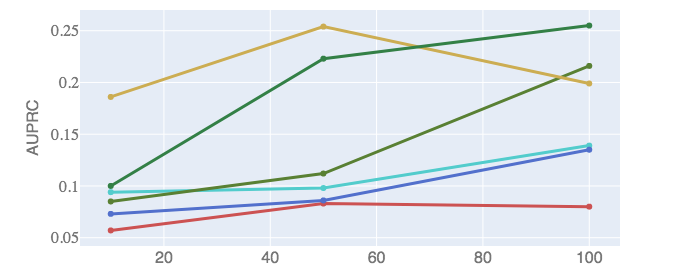}%
\label{fig:lineplot_auprc_numpatients_msseg2015}}

\caption{AUPRC of selected models trained with different numbers of healthy numbers of healthy training subjects (10, 50 and 100\%, respectively).}
\label{fig:lineplot}
\end{figure*}

In our previous experiment, we relied on 110 healthy training subjects. The question arises whether this is a sufficient amount, or if fewer scans even lead to comparable results. To give insights into the behavior of the examined models in this context, we provide a comparison of the AUPRC of conceptually most different models, all trained at varying number of healthy subjects, i.e. 10, 50 and 100\% of the available training samples. Results on the four different datasets can be seen in Fig. \ref{fig:lineplot}. The GAN-based models, which model the healthy distribution the closest due to the Wasserstein-loss, show consistent improvements in AUPRC with a growing training set. Alone the AnoVAEGAN shows a slight drop at 50\% of the training data on $\mathcal{D}_{GB}$. The overall top-performer, with one exception, is still the restoration method, here reported using the GMVAE (dense). Alone on $\mathcal{D}_{MSSEG2015}$, this GMVAE shows inconsistent behavior. Both the VAE and Context VAE, our selection from the family of VAEs with a dense bottleneck, show improved and similar performance with increasing number of training subjects on any of the MS datasets. On $\mathcal{D}_{MS}$, both models exhibit inconsistent behavior, and the VAE performs considerably better. Among all the methods, the dense AE yields the most unpredictable performance, varying greatly among different datasets and different number of healthy subjects.

% AnoVAEGAN: We relate this to improved reconstruction capabilities, providing greater attention to detail with increasing training set size, leading to fewer false positive residuals during anomaly segmentation.

\subsection{Model Complexity}

\begin{figure*}[!t]
\centering
\subfloat[MSKRI]{\includegraphics[width=0.49\textwidth]{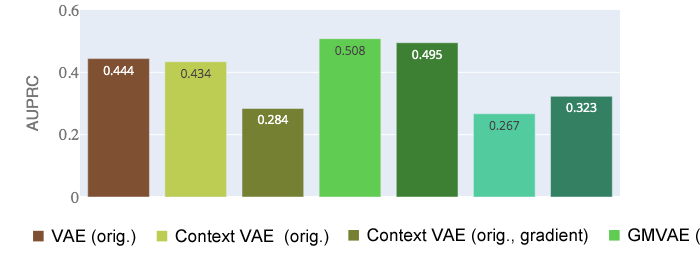}%
\label{fig:barchart_auprc_orig_mskri}}
\hfil
\subfloat[GBKRI]{\includegraphics[width=0.49\textwidth]{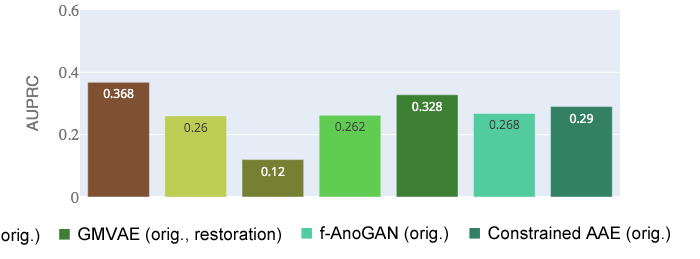}%
\label{fig:barchart_auprc_orig_gbkri}}

\subfloat[MSLUB]{\includegraphics[width=0.49\textwidth]{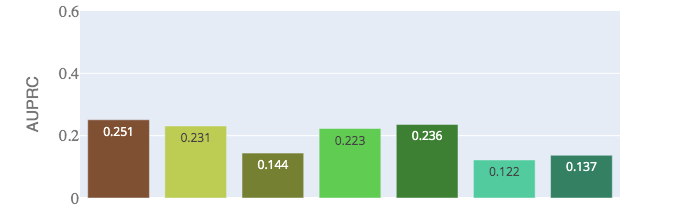}%
\label{fig:barchart_auprc_orig_mslub}}
\hfil
\subfloat[MSSEG2015]{\includegraphics[width=0.49\textwidth]{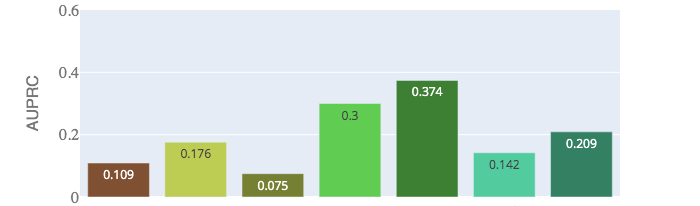}%
\label{fig:barchart_auprc_orig_msseg2015}}

\caption{AUPRC of all models and UAD approaches, using the original, more complex architectures proposed in the respective papers.}
\label{fig:orig_barchart}
\end{figure*}

To give some insights on the relation between model complexity and segmentation performance, we further rank some of the approaches based on the architectures originally proposed in the respective papers against each other. A comparison is provided on all datasets in Fig. \ref{fig:orig_barchart}. Therein, we find the VAE and the restoration-based GMVAE methods to be stable candidates. Except for $\mathcal{D}_{MSSEG2015}$, the standard VAE approach as proposed in \cite{zimmerer2018context,zimmerer2019unsupervised,baur2018deep} shows reliable performance. Similarly, the GMVAE, especially in combination with restoration-based UAD, shows good performance across all datasets. Interestingly, the more complex VAE and Context VAE models in Fig. \ref{fig:orig_barchart} show only comparable performance to the less complex models following our unified architecture (Fig. \ref{fig:barchart_auprc_unified_msseg2015}. On $\mathcal{D}_{GB}$, none of the more complex models beat the top-performing unified restoration approach. The gradient-based approach, proposed in combination with the original Context VAE, yields lower AUPRC than its unified counterpart. We relate this observation to the reconstruction capabilities of models, which improve with an increase of model parameters. With increasing complexity, larger lesions such as Glioblastoma get reconstructed better as well, which is not desirable.

%\subsection{Uncertainty}

%\todo{TODO}

%\begin{itemize}
%    \item VAE (kein dropout nötig), AEdense, AEspatial, fAnoGAN
%    \item MC=100, Dropoutrate 0.2
%    \item Pro Ansatz ein zweifarbiges Histogram mit uncertainties von anomalies vs healthy tissue
%    \item avg uncertainty für anomalien und healthy tissue
%\end{itemize}

%\subsection{Clustering in Latent Space of Healthy and Anomalous Samples}

\subsection{Reconstruction Fidelity and UAD Performance}

\begin{figure*}[!t]
\centering
\subfloat[MSKRI]{\includegraphics[width=0.49\textwidth]{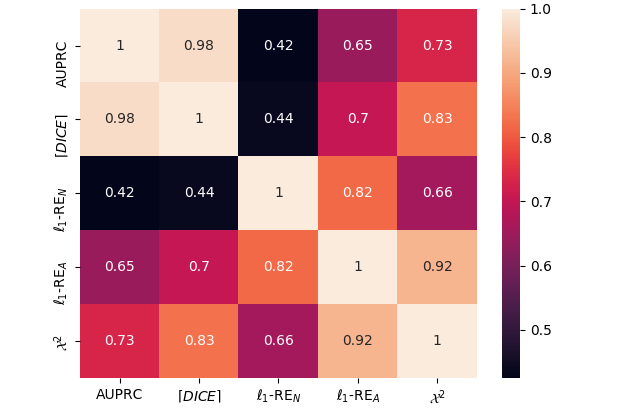}%
\label{fig:corr_mskri}}
\hfil
\subfloat[GBKRI]{\includegraphics[width=0.49\textwidth]{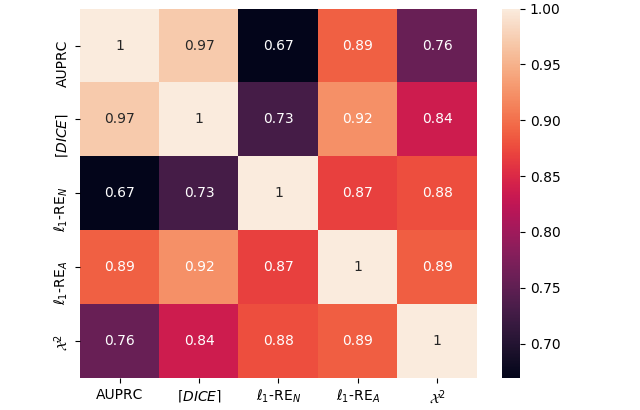}%
\label{fig:corr_gbkri}}

\subfloat[MSLUB]{\includegraphics[width=0.49\textwidth]{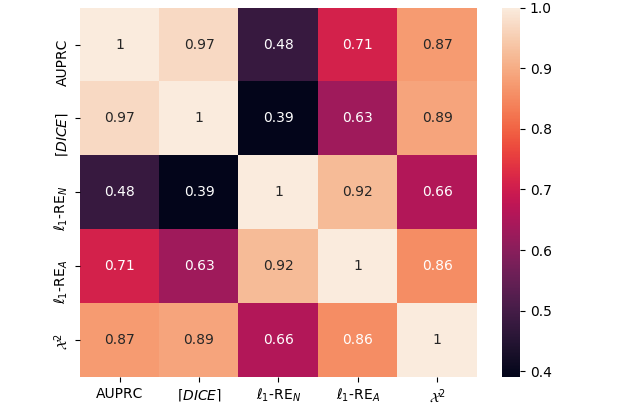}%
\label{fig:corr_mslub}}
\hfil
\subfloat[MSSEG2015]{\includegraphics[width=0.49\textwidth]{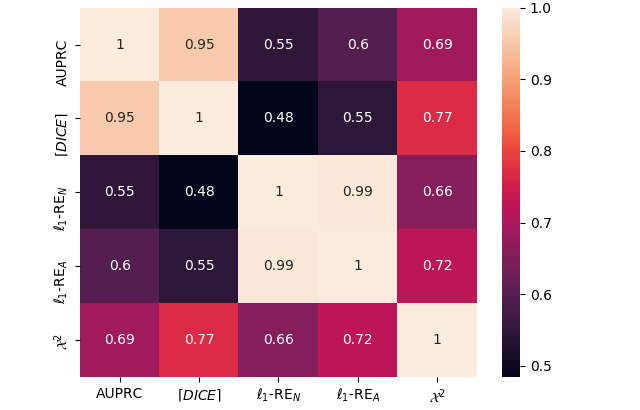}%
\label{fig:corr_msseg}}

\caption{Correlation matrices among segmentation performance, reconstruction fidelity and overlap among residual histograms of normal and anomalous intensities.}
\label{fig:corr}
\end{figure*}

From Fig. \ref{fig:examples_unified} it is clear that apart from spatial models, none of the approaches can reconstruct input perfectly, i.e. none of these methods leave healthy regions intact and substitute anomalous regions with plausible healthy anatomy. Nonetheless, some works perform better than others. We try to relate anomaly segmentation performance to the overlap between a models' residual histograms of normal and anomalous pixels and general reconstruction fidelity. Therefor, we correlate the AUPRC and $\lceil$DICE$\rceil$ to the $\mathcal{X}^2$-distance of the aforementioned histograms, and further determine how the $\mathcal{X}^2$-distance correlates with reconstruction fidelity of normal and/or anomalous tissue. We do this for every dataset separately to find out if the correlation differs across datasets and pathologies. Fig. \ref{fig:corr} shows the correlation heatmaps of aforementioned measures on all datasets.

On $\mathcal{D}_{MS}$ and $\mathcal{D}_{MSLUB}$, AUPRC and $\lceil$DICE$\rceil$ show moderate to strong correlation to the reconstruction error on anomalous pixels $\ell_1$-RE$_A$, but not so much to residuals of normal intensities $\ell_1$-RE$_N$. Their correlation to the $\mathcal{X}^2$-distance among residual histograms is the strongest. There is also a strong correlation between $\mathcal{X}^2$ and $\ell_1$-RE$_A$, the correlation to $\ell_1$-RE$_N$ is less pronounced. From these results we deduce that actual reconstruction fidelity is less important for UAD than clearly distinguishable residual histograms of normal and anomalous intensities.

For $\mathcal{D}_{GB}$, similar, but generally stronger correlations can be seen. Interestingly, there is also a moderate to strong, positive relationship between segmentation performance and magnitude of normal residuals. This indicates that with increasing reconstruction error on both normal and anomalous intensities, segmentation performance improves. We hyptothesize that models which reconstruct data well, also reconstruct tumors well. Models with generally poor reconstruction capabilities substitute tumors with poor reconstructions of healthy tissue, leading to better separability between anomalies and normal intensities.
% Reason: size and variability of tumors

On $\mathcal{D}_{MSSEG2015}$, the previously noticed correlations are hardly present. Instead, $\ell_1$-RE$_N$ and $\ell_1$-RE$_A$ are strongly correlated and seem to correlate similarly with all other metrics. This clearly reflects the poor contrast in the underlying MR images, which renders UAD unsuitable.

\subsection{Discussion}

\textbf{Ranking}|The clear winner of this comparative study is the restoration method applied to a VAE (VAE (restoration)), which achieves best performance on $\mathcal{D}_{MS}$ and $\mathcal{D}_{GB}$, i.e. works best on different pathologies, but also achieves best performance on $\mathcal{D}_{MSLUB}$, i.e. under domain shift. However, there is a downside to the restoration method, namely runtime. A restoration of a single axial slice in 500 iterations takes multiple seconds, which for an entire MR volume accumulates quickly to multiple minutes. The feed-forward nature of purely reconstruction-based approaches allows for a much faster inference. In this context, a very promising method is the reconstruction-based f-AnoGAN, which achieves best performance on the very challenging MSSEG2015 Dataset, and is only slightly inferior to the winning restoration approach on all other datasets. Also, we find that latent variable models perform better in anomaly segmentation than classic AEs. Their reconstructions tend to be more blurry, but the gap between reconstruction errors of normal and anomalous pixels is considerably higher and allows to discriminate much better between anomalies and normal tissue. Among the latent variable models, we find the VAE to be the recommended choice, as it not only performs the best, but is the easiest to optimize. It involves fewer hyperparameters than the other approaches and does not require a discriminator network, which is a critical building block in GANs. 

\textbf{Open Problems}|Despite all the recent successes of this paradigm, there are many questions yet to be answered. A key question is how to choose an Operating Point at which the continuous output i) can be binarized and a segmentation can be obtained or ii) an input sample can be considered anomalous. Most of the methods currently either rely on a held-out validation set to determine a threshold for binarization, or make use of heuristics on the intensity distribution. One such heuristic uses the 98th percentile of healthy data as a threshold, above which every value is considered an outlier \cite{baur2018deep}. It is necessary that more principled approaches for binarization are developed.

Although reconstruction fidelity here is far from perfect, the reviewed methods seem to be indeed capable of segmenting different kinds of anomalies. Nonetheless, we believe that the community should still aim for higher levels of fidelity and modeling MRI also at higher resolution to facilitate segmentation of particularly small brain lesions (e.g. MS lesions, which can become very small) and enhance precision of anomaly localization.

Another obvious downside of the reviewed methods is the necessity of a curated dataset of healthy data. It is debatable whether such methods can actually be called unsupervised or should be seen as weakly-supervised. The community should aim for methods which can be trained from all kinds of samples, even data potentially including anomalies, without the need for human ratings. You et al. \cite{you2019uad} made an initial attempt towards this direction by using a percentile-based heuristic on the training data to mask out potential outliers during training, and with so called \emph{discriminative reconstruction autoencoders}~\cite{xia2015learning} an interesting concept has recently been proposed in the Computer Vision field. All in all, more research in this direction is heavily encouraged.

Generally, the field of Deep Learning based UAD for brain imaging is rapidly growing, and without the availability of a well defined benchmark dataset the field becomes increasingly confusing. This confusion primarily arises from the different datasets used in these works, which come at different resolutions, with different lesion load and different pathologies. All of these properties make it hard to compare methods. Here, we try to give an overview of recent methods, bring them into a shared context and establish comparability among them by leveraging the same data for all approaches. Nonetheless, even the datasets used in this comparative study are limited and many open questions have to remain unanswered. Since UAD methods aim to be general, they need to be evaluated on the most representative dataset possible. Ideally, a benchmark dataset for UAD in brain MRI should comprise a vast number of healthy subjects as well as different pathologies from different scanners, covering the genders and the entire age spectrum.

To date, different works do not only employ different datasets, but also report different metrics. In addition to the benchmark, a clear set of evaluation metrics needs to be defined to facilitate comparability among methods. 

Last, the majority of approaches relies on 2D slices, but 3D offers greater opportunity and more context.

%\begin{itemize}
%    \item Choosing an Operating point (OP): how to do that in an unsupervised manner?
%    \item Evaluation metrics: Huge discrepancy
%    \item Benchmark Dataset: Do not really exists for brain MRI. In the vision field, there is the MVTec Anomaly Detection benchmark dataset, though
%    \item Resolution: Most of the works focus on low resolution data, which might be sufficient for space occupying lesions such as tumors, but not enough for small MS lesions. Baur et al.~\cite{baur2018deep} tried to address this with spatial AEs, and lately Bergmann et al in a general UAD framework~\cite{bergmann2019uninformed}
%    \item Generalization: Can AEs trained on one modality be used on other scanner data?
%    \item 2D slices vs 3D volumes: to date, most approaches operate on 2d slices, but 3D offers greater opportunity and more context (but also involves challenges)
%    \item Training from data which contains anomalies?
%\end{itemize}

%Huge variation among evaluation metrics in the papers, especially how to choose an Operating Point. A few amount of labeled data would help to choose an Operating Point.

%Discuss Restoration vs Reconstruction based methods, Spatial vs Dense Bottleneck, Variational vs Classic Autoencoders, Inpainting Autoencoders

\section{Conclusion}
In summary, we presented a thorough comparison of autoencoder-based methods for anomaly segmentation in brain MRI, which rely on modeling healthy anatomy to detect abnormal structures. We find that none of the models can perfectly reconstruct or restore healthy counterparts of potentially pathological input samples, but different approaches show different discrepancies between reconstruction-error statistics of normal and abnormal tissue, which we identify as the best indicator for good UAD performance. 

To facilitate comparability, we relied on a single unified architecture and a single image resolution. The entire code behind this comparative study, including the implementations of all methods, pre-processing and evaluation pipeline will be made publicly available and we encourage authors to contribute to it. Authors might benefit from a transparent ranking which they can report in their work without having to reinvent the wheel to run extensive comparisons against other approaches.

In our discussion, we also identify different research directions for future work. Comparing different model-complexities, their correlation with reconstruction quality and its effect on anomaly segmentation performance is another research direction orthogonal to our investigations. Determining the correlation between image resolution and UAD performance is also an open task. However, our main proposal is the creation of a benchmark dataset for UAD in brain MRI, which involves many challenges by itself, but would be very beneficial to the entire community.

% or


\section*{Acknowledgment}

The authors would like to thank their clinical partners at Klinikum rechts der Isar, Munich, for generously providing their data. S.A. was supported by the PRIME programme of the German Academic Exchange Service (DAAD) with funds from the German Federal Ministry of Education and Research (BMBF), and B. W. was supported by the DFD SFB-824 grant.

\ifCLASSOPTIONcaptionsoff
  \newpage
\fi



\bibliographystyle{IEEEtran}
\bibliography{main}

\newpage

\appendices

\section{}

%\end{IEEEbiography}


%\newpage



%\vfill


\end{document}